\documentclass[aps,prl,reprint,superscriptaddress]{revtex4-1}
\pdfoutput=1
\usepackage[version=3]{mhchem}
\usepackage[T1]{fontenc}
\usepackage{graphicx}
\usepackage{epstopdf}
\usepackage{ifpdf}
\usepackage{dcolumn} 
\usepackage{bm}
\usepackage{amssymb}
\usepackage{amsmath}
\usepackage{upgreek}
\usepackage{gensymb}
\usepackage{natmove}
\usepackage{lmodern}
\usepackage{booktabs}
\usepackage{setspace}
\usepackage{adjustbox}
\usepackage{natbib}
 \usepackage{url}
\usepackage{xcolor}
\usepackage{soul}

\begin{document}
\title{Negative Fermi-level Pinning Effect of Metal/n-GaAs(001) Junction with Graphene Interlayer}

\author{Hoon Hahn Yoon} \affiliation{Department of Physics, Ulsan National Institute of Science and Technology (UNIST), Ulsan {\sl 44919}, Republic of Korea}
\author{Wonho Song} \affiliation{Department of Physics, Ulsan National Institute of Science and Technology (UNIST), Ulsan {\sl 44919}, Republic of Korea}
\author{Sungchul Jung} \affiliation{Department of Physics, Ulsan National Institute of Science and Technology (UNIST), Ulsan {\sl 44919}, Republic of Korea}
\author{Junhyung Kim} \affiliation{School of Electrical and Computer Engineering, Ulsan National Institute of Science and Technology (UNIST), Ulsan {\sl 44919}, Republic of Korea}
\author{Kyuhyung Mo} \affiliation{Department of Physics, Ulsan National Institute of Science and Technology (UNIST), Ulsan {\sl 44919}, Republic of Korea}
\author{Gahyun Choi} \affiliation{Department of Physics, Ulsan National Institute of Science and Technology (UNIST), Ulsan {\sl 44919}, Republic of Korea}
\author{Hu Young Jeong} \affiliation{UNIST Central Research Facilities (UCRF), Ulsan National Institute of Science and Technology (UNIST), Ulsan {\sl 44919}, Republic of Korea}
\author{Jong Hoon Lee} \affiliation{UNIST Central Research Facilities (UCRF), Ulsan National Institute of Science and Technology (UNIST), Ulsan {\sl 44919}, Republic of Korea}
\author{Kibog Park} \email{kibogpark@unist.ac.kr} \affiliation{Department of Physics, Ulsan National Institute of Science and Technology (UNIST), Ulsan {\sl 44919}, Republic of Korea} \affiliation{School of Electrical and Computer Engineering, Ulsan National Institute of Science and Technology (UNIST), Ulsan {\sl 44919}, Republic of Korea}
 
\begin{abstract}
It is demonstrated that the electric dipole layer due to the overlapping of electron wavefunctions at metal/graphene contact results in negative Fermi-level pinning effect on the region of GaAs surface with low interface-trap density in metal/graphene/n-GaAs(001) junction. The graphene interlayer takes a role of diffusion barrier preventing the atomic intermixing at interface and preserving the low interface-trap density region. The negative Fermi-level pinning effect is supported by the Schottky barrier decreasing as metal work-function increasing. Our work shows that the graphene interlayer can invert the effective work-function of metal between $high$ and $low$, making it possible to form both Schottky and Ohmic-like contacts with identical (particularly $high$ work-function) metal electrodes on a semiconductor substrate possessing low surface-state density.
\end{abstract}

\maketitle

\section{I. INTRODUCTION}
The interfacial physics and chemistry are key elements in understanding the electrostatic environment and the associated carrier transport at metal/graphene interface. The change of electrostatic potential across metal/graphene interface is determined by not only the electron transfer between metal and graphene caused by the difference in work-function but also the formation of an electric dipole layer originating from the off-centric distribution of interacting electrons at the interface \cite{DopingGraphene1,DopingGraphene2,DopingGraphene3}. The electrical properties of metal/graphene and graphene/semiconductor contacts have been studied actively as separate physical systems \cite{GrapheneContacts1,GrapheneContacts2}. However, the charge carrier transport across metal/graphene/semiconductor junction as a whole, particularly taking into account the interaction dipole charges at metal/graphene contact, has not received proper attention. The interests in metal/graphene/semiconductor junction stem mainly from the possibility of modulating the electron energy barrier at interface with the graphene interlayer \cite{InterfaceGraphene1,InterfaceGraphene2,InterfaceGraphene3,InterfaceGraphene4,InterfaceGraphene5}. Recently, it has been reported that the atomically-impermeable and electronically-transparent properties of graphene inserted at metal/n-Si(001) junctions can be used to form an intact Schottky contact and investigate the interface Fermi-level pinning effect \cite{InterfaceGraphene1}. The graphene interlayer prevents the interface intermixing of metal and semiconductor atoms which occurs on the region of Si surface with very thin or no native oxide layers and leads to the formation of small-area patches with low local Schottky barriers. In this study, the graphene layer inserted at metal/n-GaAs(001) junction is found to bring about a different phenomenon although it still plays a role of diffusion barrier. It is known that the reconstructed surface of III-V compound semiconductor contains generally a low density of surface states within its band gap unless the surface structure gets compromised by some extrinsic factors including defects, oxidation, and atomic intermixing with other materials \cite{SurfacesInterfacesofElectronicMaterials}. Accordingly, the regions on the surface of our GaAs(001) substrate with very thin or no native oxide layer is considered also to possess a relatively low density of surface states. However, the material intermixing on those low-surface-state regions, occurring during metal deposition, can induce a substantial amount of interface-trap states similarly to the region of GaAs surface with normal native oxides. Then, the entire region of metal/n-GaAs(001) junction will be under relatively strong interface Fermi-level pinning effect. On the other hand, if a graphene layer is inserted at metal/n-GaAs(001) interface, the material intermixing on the regions with very thin or no native oxide layer no longer occurs and thus those regions can stay with a small density of interface-trap states, bearing pretty weak Fermi-level pinning \cite{LowDit1,LowDit2,LowDit3,LowDit4,LowDit5,LowDit6}. From the growth mechanism of native oxide, the regions with low interface-trap density will be small in size and randomly scattered on the GaAs surface. Due to the strong Fermi-level pinning on the prevailing surrounding area, the metal-dependent variation of current-voltage (I-V) characteristics of metal/graphene/GaAs junction is expected to originate from the small patches of low interface-trap density \cite{TungPaper, PC}.

\begin{figure*}[!t]
\includegraphics{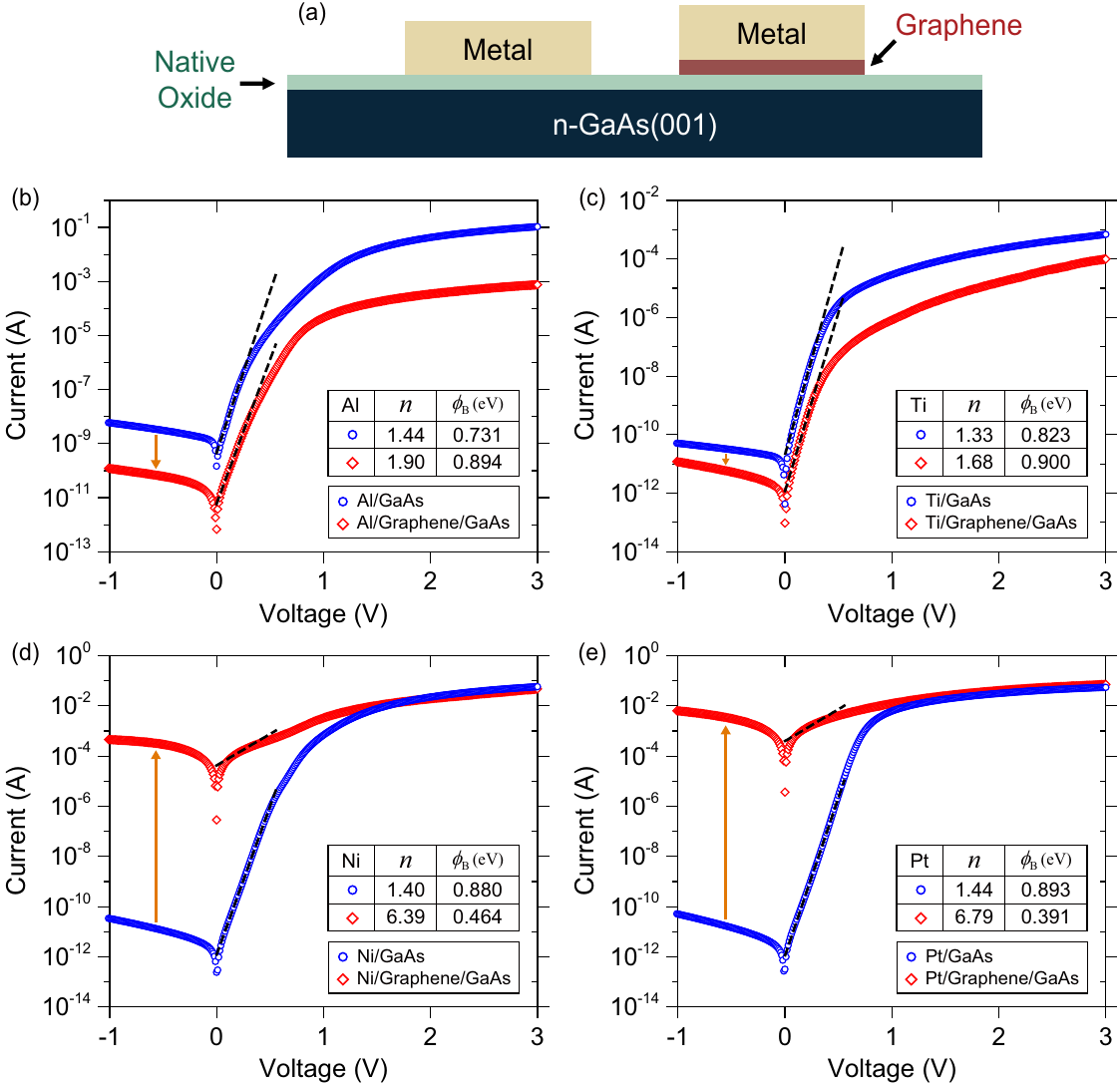}
\caption{\label{fig:1}Device schematic and I-V measurements. (a) Schematic of metal/GaAs with and without a graphene interlayer. (b)-(e) I-V curves measured on the metal/GaAs and metal/graphene/GaAs junctions with Al (b), Ti (c), Ni (d), and Pt (e) electrodes. Averaged over several different junctions, the Schottky barrier $\phi_{B}$ and ideality factor $n$ for each metal are listed in the table inset.}
\end{figure*}

\section{II. EXPERIMENTAL SECTION}
\subsection{A. Device Fabrication}
The metal/GaAs and metal/graphene/GaAs junctions were prepared as follows. The Si-doped n-type ($N_{D}$ $\simeq$ 5$\times$$10^{16}$ cm$^{-3}$) GaAs wafer grown by the vertical gradient freeze (VGF) method was purchased from the MTI Corporation. The GaAs substrate was first etched in a 1:1 NH$_{4}$OH:H$_{2}$O solution for 1 min followed by DI water rinsing \cite{Wet}. After that, Ti/Pt/Au (50/200/5000 {\AA}) ohmic contacts were formed near the edge of GaAs substrate by using e-beam evaporation and rapid thermal annealing at 400 $^\circ$C for 300 sec \cite{Ohmic}. The sample was treated in a 1:1 NH$_{4}$OH:H$_{2}$O etching solution for 3 min followed by DI water rinsing again right before the graphene transfer. The monolayer graphene synthesized on a Cu foil with chemical vapor deposition (CVD) was purchased from the Graphene Supermarket. In order to minimize the trapping of water molecules between graphene and GaAs surface and the residues during the transfer process \cite{Semi-Dry Transfer}, we used the dry transfer method adopting thermal release tapes. The graphene layer was partially transferred on the GaAs surface in order to form metal/GaAs and metal/graphene/GaAs junctions simultaneously on the same GaAs substrate. The structural qualities of transferred graphene were examined with Raman spectrum measurements (see Supporting Information for the Raman spectrum measurements) \cite{Raman}. After the graphene transfer, circular metal electrodes (Al, Ti, Ni, and Pt) of average diameter $\sim$500 $\mu$m were deposited through a shadow mask on the sample surface by using e-beam evaporation. The metal/GaAs junctions were formed on the graphene-uncovered area. Finally, the graphene uncovered by metal electrodes was removed with reactive ion etching (RIE) to isolate each junction.

\begin{figure*}[!t]
\includegraphics[width=16cm]{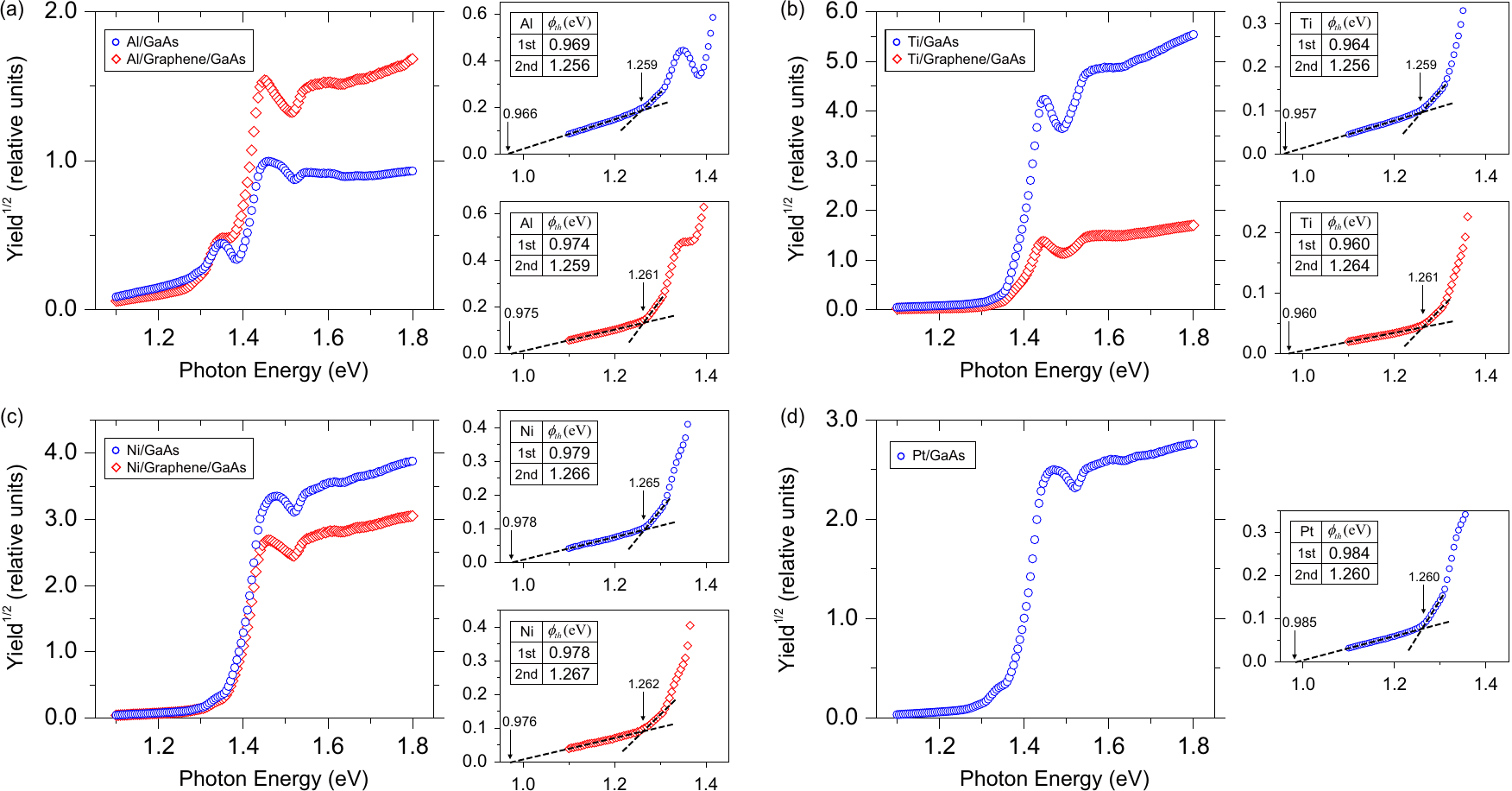}
\caption{\label{fig:2}IPE measurements. (a)-(d) Square root of IPE quantum yield as a function of photon energy measured on the metal/GaAs and metal/graphene/GaAs junctions with Al (a), Ti (b), Ni (c), and Pt (d) electrodes. The magnified views of the threshold region with linear extrapolation are also shown. Averaged over several different junctions, two IPE thresholds $\phi_{th}$ are listed in the table inset of magnified view.}
\end{figure*}

\subsection{B. Current-Voltage Measurement}
The schematic of metal/GaAs junctions with and without a graphene interlayer is illustrated in Figure~\ref{fig:1}a. Figure~\ref{fig:1}b-e show the I-V characteristics of metal/GaAs and metal/graphene/GaAs junctions. As shown in Figure~\ref{fig:1}b,c, the Al/graphene/GaAs and Ti/graphene/GaAs junctions possess the typical rectifying characteristics. In fact, their reverse bias currents are somewhat smaller (more rectifying) than the Al/GaAs and Ti/GaAs junctions. On the other hand, the reverse bias currents of Ni/graphene/GaAs and Pt/graphene/GaAs junctions are significantly larger than those of Ni/GaAs and Pt/GaAs junctions (Figure~\ref{fig:1}d,e). The rectifying behaviors of both junctions have almost disappeared, turning into Ohmic-like ones. This is quite counter-intuitive since the local Schottky barrier formed on the region of GaAs surface with low interface-trap density is expected to be low for Al/graphene/GaAs and Ti/graphene/GaAs junctions and high for Ni/graphene/GaAs and Pt/graphene/GaAs junctions when considering the difference in metal work-function.

\begin{figure*}[!t]
\includegraphics{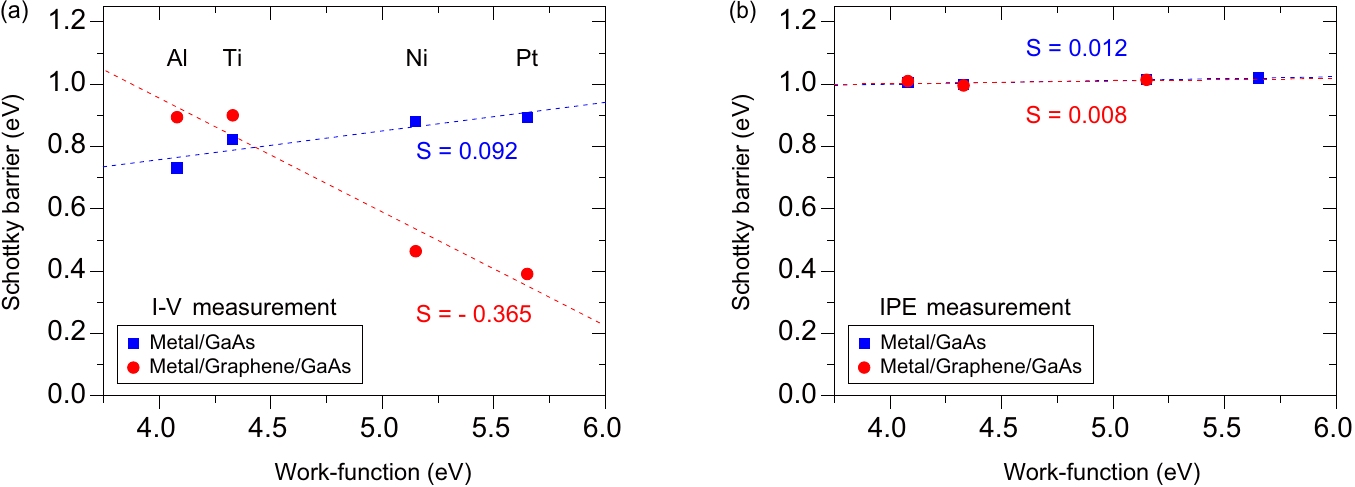}
\caption{\label{fig:3}Measured pinning strength. I-V extracted ({\bf a}) and IPE extracted ({\bf b}) Schottky barriers of metal/GaAs and metal/graphene/GaAs junctions as a function of metal work-function. The extracted pinning factor $S = \partial \phi_{B}/\partial \phi_{M}$ is labeled for each case.}
\end{figure*}

\subsection{C. Internal Photoemission Measurement}
The Internal Photoemission (IPE) yield spectra \cite{InterfaceGraphene1,IPE,AuGaAsIPE} for metal/GaAs and metal/graphene/GaAs junctions are shown in Figure~\ref{fig:2}. Since the IPE measurement reveals the motion of ballistic electron across the junction interface, its signals will be contributed collectively by all regions in the junction. However, if a junction is mostly uniform and only a small areal fraction of isolated patches exists with their interface energy barriers different from those on the uniform surrounding region, the IPE spectrum will be dominated by the prevailing region and the contribution from the small patches will be almost invisible. In the IPE spectra of metal/GaAs junctions, two thresholds $\phi_{th}$ for the conduction band minima ($\Gamma$ and $L$ valleys) of GaAs \cite{Valley} are clearly observed. The first threshold represents the commonly accepted Schottky barrier of metal/GaAs junction, which reflects the electron transmission from metal into the $\Gamma$ valley in GaAs. The additional transmission into the $L$ valley in GaAs corresponds to the second threshold. The gap between first and second thresholds is about $\sim$0.29 eV, which is in excellent agreement with the known energy separation between $\Gamma$ and $L$ valleys. However, the observation of additional transmission into the $X$ valley of GaAs is disturbed by the humps of IPE yield around 1.33 eV \cite{Localized} (see Supporting Information for details on the direct optical excitation from localized interface states). Since the two thresholds are extracted to be pretty much identical for all four metal electrodes (Al, Ti, Ni, Pt) and no signature for additional threshold is observed, the Fermi-level pinning can be concluded to be strong uniformly throughout the entire metal/GaAs junction as known well. The metal/graphene/GaAs junctions also show two common thresholds in their IPE spectra for Al, Ti, and Ni electrodes which are quite similar to those for the metal/GaAs junctions. In case of Pt/graphene/GaAs junction, the large leakage current, confirmed in the I-V curve (Figure~\ref{fig:1}e), overwhelmed the photocurrent completely so that we could not determine the corresponding IPE yield at all (Figure~\ref{fig:2}d). The similarity in the IPE thresholds of metal/graphene/GaAs junction for different metal electrodes implies that the Fermi-level pinning is strong in the vast majority of junction area where the regular oxide layer is expected to reside and bear high interface-trap density. Here, there are several things to note about characterizing the fabricated junctions. All transport measurements including I-V and IPE measurements were carried out at room temperature. The power of incident light was measured separately by using a photodiode, HAMAMATSU S2281-04 (Si photodiode for the wavelength of 200-1180 nm) for acquiring the quantum yield of IPE\cite{InterfaceGraphene1,IPE,AuGaAsIPE}. After the measurements of electrical properties, the layer structures of fabricated junctions were examined by taking transmission electron microscope (TEM) images (see Supporting Information for the TEM images) \cite{TEM,Oxide1,Oxide2}.

\begin{figure*}[!t]
\includegraphics[width=16.5cm]{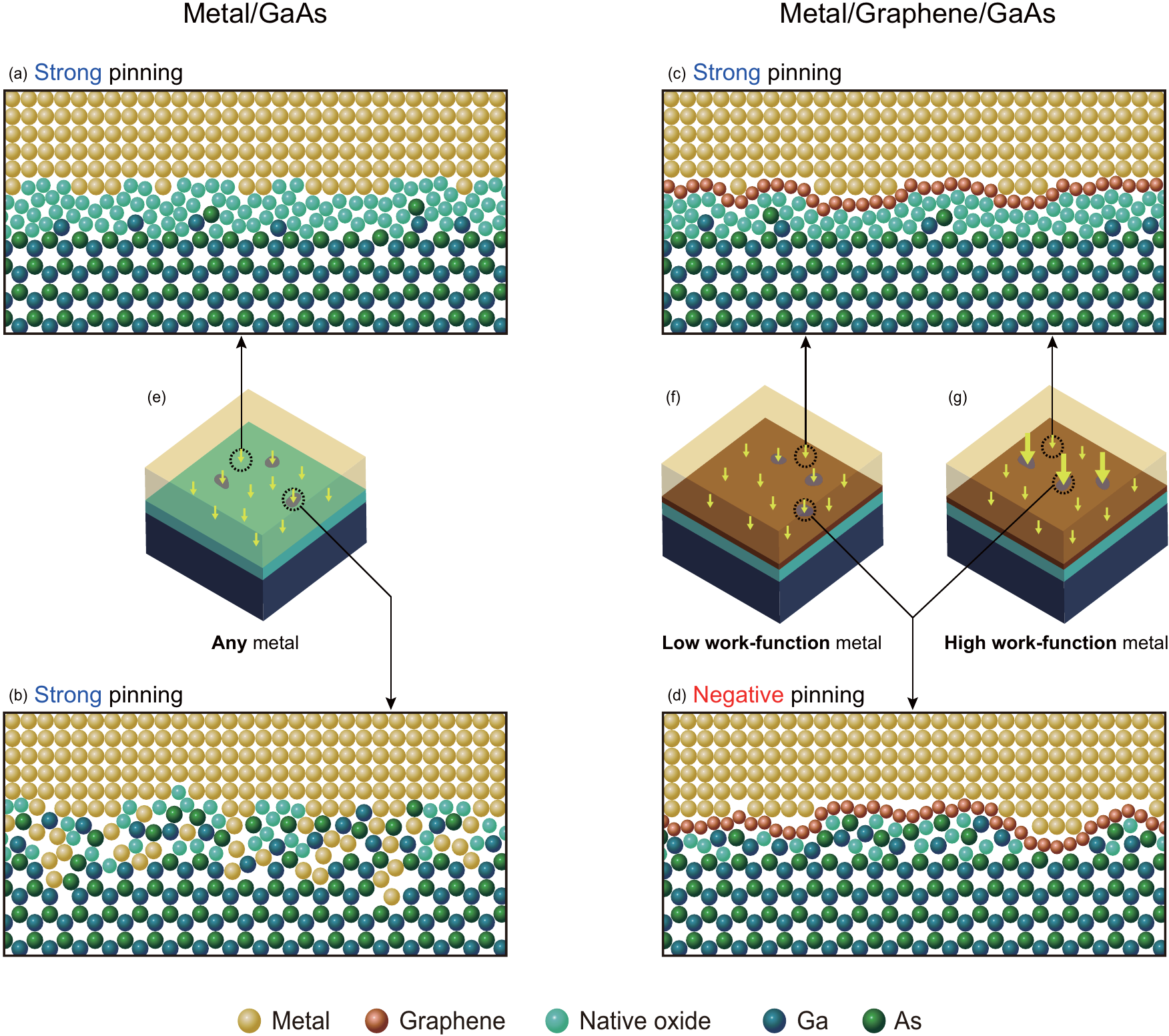}
\caption{\label{fig:4}Atomic arrangement at interface. (a)-(d) Cross-sectional schematics of atomic arrangements at the interfaces of metal/GaAs and metal/graphene/GaAs junctions reflecting the spatial-inhomogeneity of native oxide layer. (e)-(g) Schematics of metal/GaAs and metal/graphene/GaAs junctions with low and high metal work-functions. The black dashed circles in (e)-(g) are linked to the corresponding atomic arrangements at interfaces in (a)-(d). The size of yellow arrows indicates the relative magnitude of junction current.}
\end{figure*}

\section{III. RESULTS AND DISCUSSION}
\subsection{A. Negative Fermi-level Pinning Effect}

Averaged over several different junctions, the Schottky barriers of metal/GaAs and metal/ graphene/GaAs junctions were obtained from the I-V and IPE measurements. They were determined by either fitting the measured I-V curves shown in Figure~\ref{fig:1} to the thermionic-emission theory \cite{Thermionic} or extracting the first threshold of IPE yield shown in Figure~\ref{fig:2} with the image force lowering included. Figure~\ref{fig:3}a,b show the obtained Schottky barriers $\phi_{B}$ of metal/GaAs and metal/graphene/GaAs junctions as a function of metal work-function $\phi_{M}$. From the I-V measurements, the pinning factor $S$, defined to be $\partial \phi_{B}/\partial \phi_{M}$ \cite{Pinning,SzeText,SzePaper}, are obtained to be 0.092 and -0.365 for metal/GaAs and metal/graphene/GaAs junctions, respectively. As mentioned previously, it is quite interesting that the Schottky barrier of metal/graphene/GaAs junction decreases as the metal work-function increases, reflected on the negative value (-0.365) of $S$. In normal circumstances, the pinning factor ranges between 0 and 1 depending on the pinning strength. In case of IPE measurements, the Schottky barriers for metal/GaAs and metal/graphene/GaAs junctions are extracted to be very similar to each other for all the metals except for Pt. Just as a reminder, the IPE spectrum for Pt/graphene/GaAs couldn't be obtained due to the large junction leakage. The pinning factors in the IPE measurements are found to be 0.012 and 0.008 for metal/GaAs and metal/graphene/GaAs junctions, respectively, both indicating strong Fermi-level pinning effect.

Based on the I-V and IPE measurements described above, it seems plausible to conclude the following spatial distributions for the interface Fermi-level pinning and associated current flows in metal/GaAs and metal/graphene/GaAs junctions. For metal/GaAs junctions, as shown in Figure~\ref{fig:4}a,b, the entire region of GaAs surface will have high interface-trap density due to oxidation (native oxide, Figure~\ref{fig:4}a) or material intermixing with metal atoms (Figure~\ref{fig:4}b) \cite{LowDit1,LowDit2,LowDit3,LowDit4,LowDit5,LowDit6}. Hence, the interface Fermi-level pinning effect will appear to be strong throughout the entire region of junction and the junction current will be distributed uniformly (Figure~\ref{fig:4}e). In case of metal/graphene/GaAs junctions, the prevailing region with regular native oxides will have a large density of interface-trap states (Figure~\ref{fig:4}c) and the Fermi-level pinning on this region will be strong similarly to the metal/GaAs junction case.  Meantime, the small patches with very thin or no native oxide layer will maintain low interface-trap density thanks to the protection of graphene interlayer (Figure~\ref{fig:4}d) and the observed negative Fermi-level pinning will occur here. Due to the negative pinning, these small patches will have current flows comparable to the surrounding regions for $low$ work-function metal electrodes (Figure~\ref{fig:4}f) forming relatively high interface energy barriers. On the other hand, the current flow through the small patches will be a lot more in comparison with the surrounding regions for $high$ work-function metal electrodes (Figure~\ref{fig:4}g) bearing leaky current paths with low interface energy barriers.

\subsection{B. Calculated energy band profile across junction}

\begin{figure*}[!t]
\includegraphics[width=16.5cm]{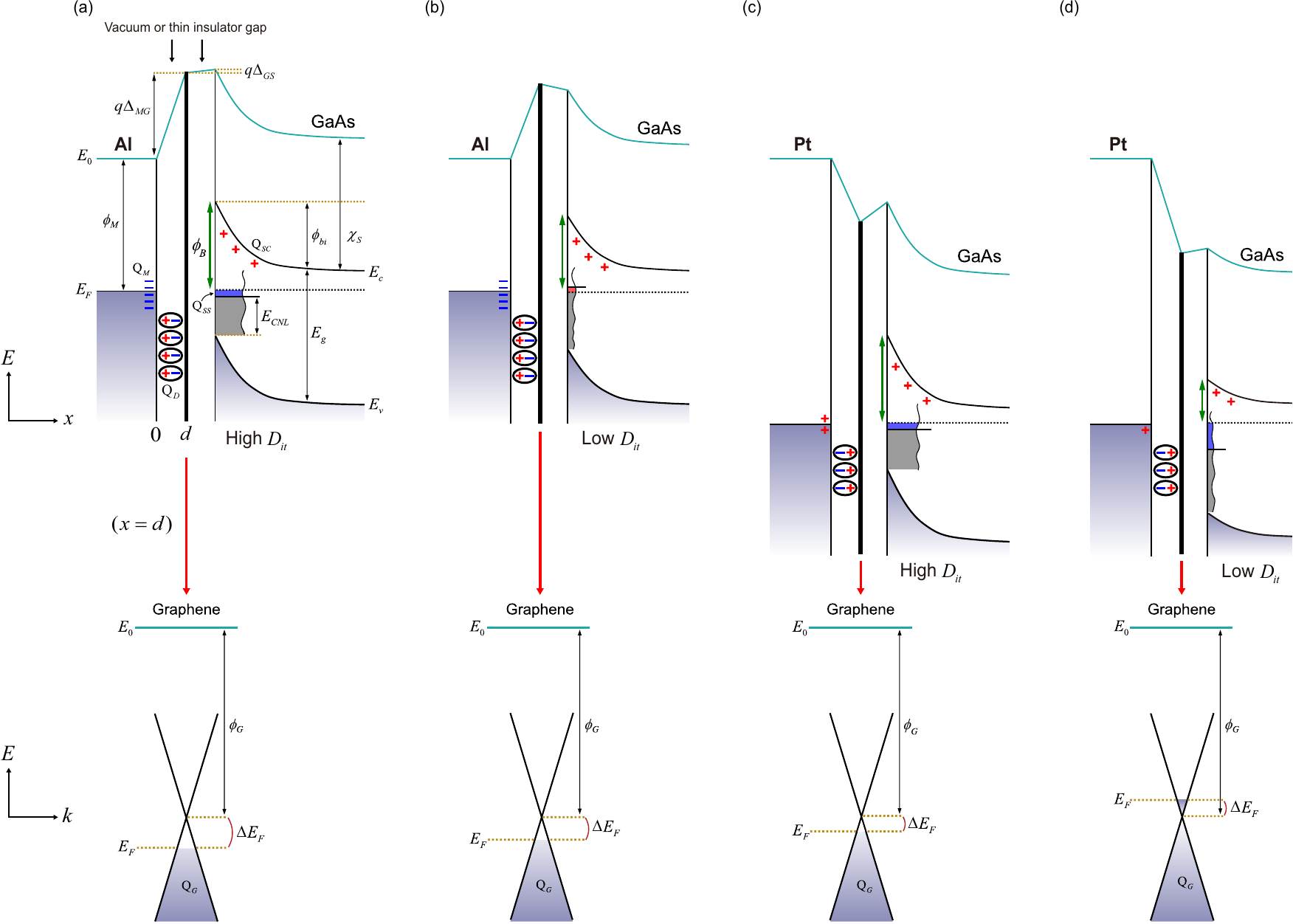}
\caption{\label{fig:5}Energy band alignments. (a)-(d) Energy band alignments of Al/graphene/GaAs and Pt/graphene/GaAs junctions for high (a, c) and low (b, d) interface-trap density ($D_{it}$) regions. Each of the energy band profiles shown in the upper side is drawn as a function of position in the junction (the metal surface is set to be $x = 0$) and the corresponding linear energy dispersion relation of graphene with the Fermi-level indicated is drawn below each band profile. $\Delta_{MG}$: Potential change across metal/graphene interface, $\Delta_{GS}$: Potential change across graphene/GaAs interface, $E_{0}$: Vacuum level, $E_{F}$: Metal Fermi-level, $E_{c}$: Conduction band edge, $E_{v}$: Valence band edge. $\phi_{M}$: Metal work-function, $\phi_{B}$: Schottky barrier, $\phi_{bi}$: Built-in potential, $\chi_{S}$: Electron affinity of GaAs, $E_{g}$: Band gap of GaAs, $E_{CNL}$: Charge neutrality level of GaAs, $Q_{M}$: Free charge density on metal surface, $Q_{D}$: Interaction dipole charge density at metal/graphene contact, $Q_{SS}$: interface-trap charge density on GaAs surface, $Q_{SC}$: Space charge density in the depletion region of GaAs substrate, $\Delta$$E_{F}$: Fermi-level shift of graphene, $\phi_{G}$:  Graphene work-function, and $Q_{G}$: Doping charge density of graphene.}
\end{figure*}

In order to understand the unusual negative Fermi-level pinning in the metal/graphene/GaAs junction, the electron energy band profiles across the junction were obtained by performing finite element electrostatic modeling with a commercial software package FlexPDE \cite{PDE1,PDE2,PDE3} (see Supporting Information for details on the finite element electrostatic modeling). All parameters were implemented with the values measured or reported in the literature \cite{SzeText,GaAsDit,CRC,MP,Work-function}. The key variable in our modeling is the interaction dipole charge density ($Q_{D}$) at metal/graphene contact which was supported by density functional theory (DFT) calculations to exist due to the off-centric distribution of the overlapped electron wavefunctions in the gap between metal and graphene layer \cite{DopingGraphene1,DopingGraphene2}. Here, we note that $Q_{D}$ indicates the charges of interaction dipole layer on the graphene side. First of all, we have calculated the Schottky barrier ($\phi_{B}$) and pinning strength ($S$) for various interface-trap density ($D_{it}$) on GaAs surface by assuming $Q_{D}$ = 0 in order to verify whether or not the observed negative Fermi-level pinning can occur without the interaction dipole layer.  If we do not consider the interaction dipole layer, the pinning strength is always calculated to be positive for $10^{12}$ eV$^{-1}$$\cdot$cm$^{-2} \leq D_{it} \leq 10^{15}$ eV$^{-1}$$\cdot$cm$^{-2}$ which is the typical range for III-V compound semiconductors \cite{GaAsDit} (see Table~S1 in Supporting Information). In addition, if we estimate the $Q_{D}$ to match with the I-V measured $\phi_{B}$ for $D_{it} \geq 10^{13}$ eV$^{-1}$$\cdot$cm$^{-2}$, the $Q_{D}$ increases dramatically to have the physically impossible values (see Table~S2 in Supporting Information). This confirms that the negative Fermi-level pinning effect observed in the I-V measurements should occur on the region with low interface-trap density ($D_{it} \leq 10^{13}$ eV$^{-1}$$\cdot$cm$^{-2}$) as concluded from the IPE measurements. Since the interaction dipole layer at metal/graphene contact is expected to be uniformly formed throughout the entire contact, it is reasonable to assume that each metal will have a certain constant $Q_{D}$ in metal/graphene/GaAs junction regardless of the locally-varying $D_{it}$ on GaAs surface. In the electrostatic modeling, we adopted the $Q_{D}$ obtained for $D_{it} = 5\times10^{12}$ eV$^{-1}$ cm$^{-2}$ to fit the I-V measured $\phi_{B}$ and $D_{it} = 5\times10^{14}$ eV$^{-1}$ cm$^{-2}$ was used for the prevailing region with high interface-trap density. The calculated relevant potentials and charges are listed (see Table~S3 in Supporting Information) and the band profiles of Al/graphene/GaAs and Pt/graphene/GaAs junctions are shown in Figure~\ref{fig:5} as the representatives of low and high work-function metal electrodes.

\begin{figure*}[!t]
\includegraphics{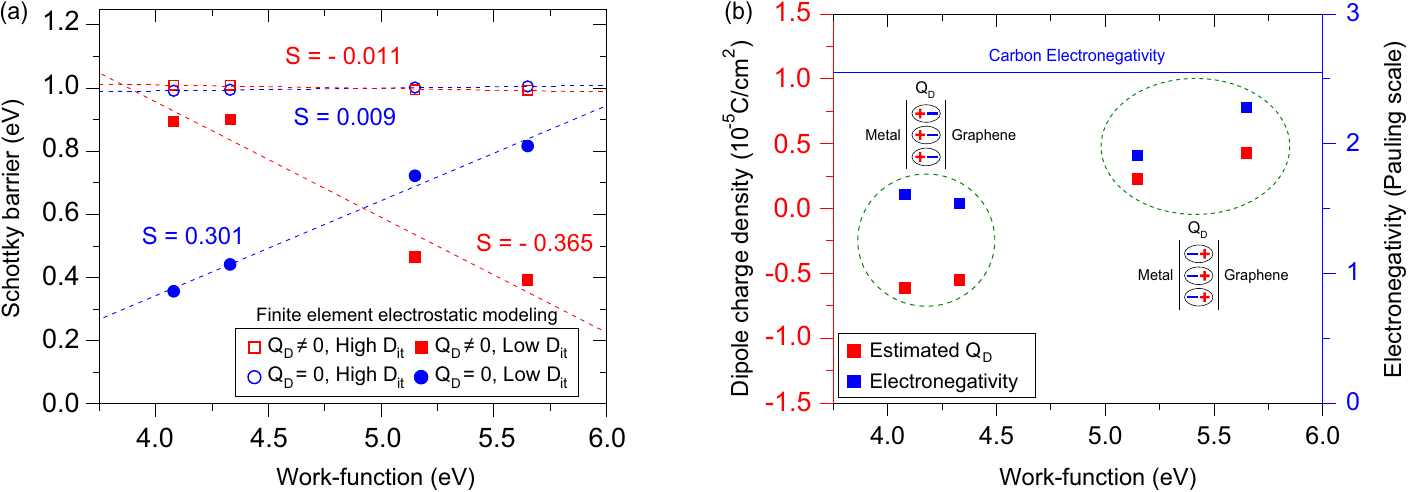}
\caption{\label{fig:6}Calculated pinning strength, correlation of electronegativity and interaction dipole charge. (a) The Schottky barrier of metal/graphene/GaAs junction obtained from the electrostatic modeling as a function of metal work-function for high ($5\times10^{14}$ eV$^{-1}$ cm$^{-2}$) and low ($5\times10^{12}$ eV$^{-1}$ cm$^{-2}$) $D_{it}$. The modeling is performed for both $Q_{D} = 0$ and $Q_{D} \neq 0$. Here, $D_{it}$ is interface-trap density on GaAs surface and $Q_{D}$ is the interaction dipole charge density at metal/graphene contact. The extracted pinning factor $S = \partial \phi_{B}/\partial \phi_{M}$ for each case is also labeled. (b) The $Q_{D}$ obtained from the electrostatic modeling and the electronegativity of metal atom reported in the previous study \cite{Electronegativity} as a function of metal work-function. The green dashed circles indicate two distinct groups with opposite polarities of $Q_{D}$ and the corresponding interaction dipole configurations are also shown.}
\end{figure*}

\subsection{C. Electronegativity difference and interaction dipole charge}
Figure~\ref{fig:6}a shows the Schottky barriers extracted from the modeling for four different combinations of $Q_{D}$ and $D_{it}$. As seen in the figure, the negative pinning factor for metal/graphene/ GaAs junction is obtained only with non-zero $Q_{D}$ for low $D_{it}$. In case of Pt/graphene/GaAs junction, it is found that $Q_{D}$ should be positive to obtain the Schottky barrier measured to be small. This polarity actually coincides with the DFT calculation results mentioned previously \cite{DopingGraphene1,DopingGraphene2}. Meanwhile, $Q_{D}$ for Al/graphene/GaAs junction is found to be negative to match with the measured Schottky barrier. According to the DFT calculations, $Q_{D}$ for the Al/graphene/GaAs junction will still be positive although its magnitude becomes much smaller than that of Pt/graphene/GaAs junction. However, the positive $Q_{D}$ will lower the Schottky barrier always, relative to the junction without the graphene interlayer, since the electrostatic potential drops across the gap between metal and graphene layer. Our electrostatic modeling indicates clearly that the Schottky barrier of Al/graphene/GaAs junction is quite low with zero or positive $Q_{D}$. Thus it seems inevitable for $Q_{D}$ to be negative for bearing the experimentally-measured large Schottky barrier.

It is known well that the exchange correlation potential for electrons is attractive in front of a metal surface and the electrons of an atom adsorbed on the metal surface is attracted toward the metal side \cite{Exchange-Correlation}. Hence, the positive $Q_{D}$ is quite understandable in terms of this exchange correlation. Then, the question is what induces the negative $Q_{D}$ of Al/graphene/ GaAs junction. One possibility that we can think of is the electronegativity difference between metal and carbon atom. Figure~\ref{fig:6}b shows the $Q_{D}$ obtained from the electrostatic modeling and the electronegativity \cite{Electronegativity} of a metal atom as a function of metal work-function. As shown in the figure, the electronegativities of all the metals used in our work are smaller than that of carbon. This implies that the interacting electrons between metal and graphene will be attracted more toward the graphene side. Especially, the electronegativities of Al and Ti with relatively low work-functions are quite small in comparison with carbon. Accordingly, the driving force shifting the interacting electrons toward the graphene side can be significant, overcoming the tendency of shifting them toward the metal side due to the exchange correlation and leading to the negative $Q_{D}$. For Ni and Pt with relatively high work-functions, the difference of electronegativity from carbon becomes smaller. Therefore, in this case, the interacting electrons are expected to be shifted toward the metal by more prominent exchange correlation so that the positive $Q_{D}$ is induced. Here, it is noted that the mechanism described above is only a possibility relying on the two known factors of exchange correlation and atomic electronegativity which have been addressed independently. One point consistent between the DFT calculation and our electrostatic modeling is that the tendency of shifting the interacting electrons toward the metal side is reduced significantly for low work-function metals. However, what is apparent from the fundamental physics standpoint, supported from the electrostatic modeling, is that the sign change of $Q_{D}$ is the essential requirement for the negative Fermi-level pinning observed in our experiments.

\section{IV. CONCLUSIONS}
In conclusion, we report the negative Fermi-level pinning effect observed experimentally in metal/graphene/n-GaAs(001) junction. The low interface-trap density regions, protected by the graphene interlayer, are found to have the local Schottky barrier affected directly by the interaction dipole layer at metal/graphene contact. In order to explain the observed negative Fermi-level pinning, it is found that the interacting electrons at the metal/graphene interface should be attracted more toward the graphene side for low work-function metals, bearing the increase of electrostatic potential across the interface. For high work-function metals, they should be attracted more toward the metal side, making the electrostatic potential decrease across the interface accordingly. Based on our work, it can be claimed that the graphene interlayer can invert the effective work-function of metal between high and low, making it possible to form both Schottky and Ohmic-like contacts with identical (particularly $high$ work-function) metal electrodes on a semiconductor substrate possessing low surface-state density.

\section{Author Contributions}
H.H.Y and K.P. conceived and designed the experiments. H.H.Y., J.K., and K.M. fabricated the devices. H.H.Y. and W.S. performed the measurements. H.H.Y., W.S., S.J., J.K., K.M., G.C., and K.P. analyzed the measured data. W.S. and S.J. performed the electrostatic modeling. H.Y.J. and J.H.L. took TEM images. All authors discussed the results and contributed to writing the manuscript. H.H.Y. and W.S. contributed equally to this work.

\section{Acknowledgements}
This work was supported by Space Core Technology Development Program (2016M1A3A3A02017648), Basic Science Research Program (2016R1A2B4014762, 2019R1F1A1057767), and Global Ph.D Fellowship Program (2015H1A2A1033714) through the National Research Foundation funded from the Ministry of Science and ICT in Korea. This work has also benefited from the use of the facilities at UNIST Central Research Facilities.

\end{document}

\subsection{}
~

\section{Direct Optical Excitation from Localized Interface States}
\begin{figure*}[!]
\sfigone
\includegraphics{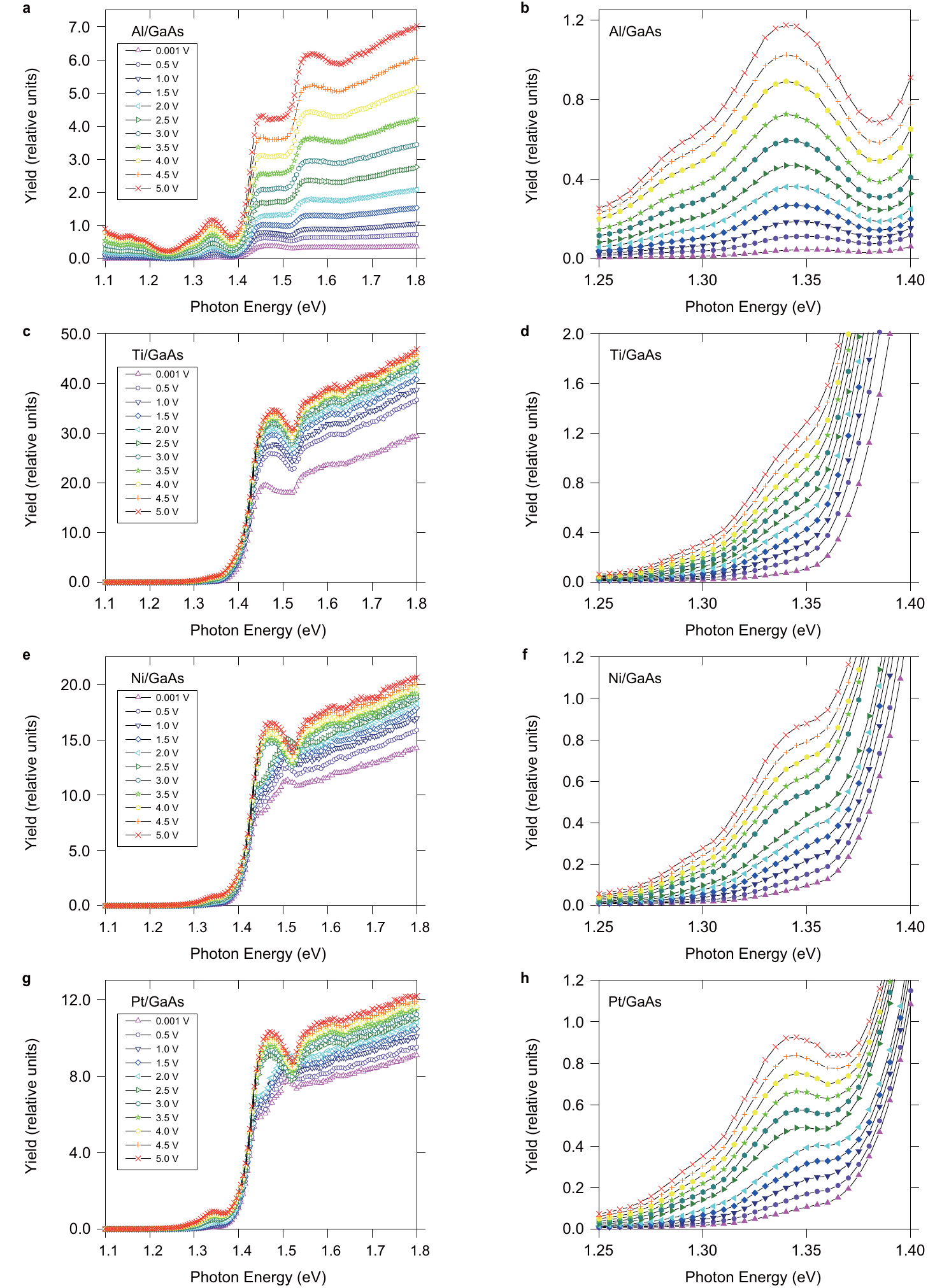}
\caption{\label{fig:s1}IPE quantum yield as a function of photon energy measured on the metal/GaAs junctions with Al (a), Ti (c), Ni (e), and Pt (g) electrodes under applied reverse bias varying from 0.001 to 5 V. The magnified views of the curves around 1.33 eV are shown in (b), (d), (f), and (h).}
\end{figure*}

~~
The IPE yield spectra for metal/GaAs and metal/graphene/ GaAs junctions are shown in Figure~2 \cite{IPE,AuGaAsIPE}. As pointed out in the manuscript, the IPE signals will be mainly from the prevailing regions of GaAs surface with high interface-trap density. Two IPE thresholds $\phi_{th}$ are clearly observed for the conduction band minima ($\Gamma$ and $L$ valleys) in GaAs \cite{Valley}. The first threshold represents the commonly accepted Schottky barrier of metal/GaAs junction, which reflects the electron transmission from metal into the $\Gamma$ valley in GaAs. The additional transmission into the $L$ valley in GaAs corresponds to the second threshold. The gap between first and second thresholds is about 0.29 eV, which is in excellent agreement with the known energy separation between $\Gamma$ and $L$ valleys. However, the observation of additional transmission into the $X$ valley in GaAs is disturbed by the humps of IPE yield around 1.33 eV, which are considered to be caused by the direct optical excitation from the localized interface states below the charge neutrality level to the $\Gamma$ or $L$ valleys. These localized interface states responsible for the strong Fermi-level pinning at metal/GaAs contact are filled with electrons since they are below the charge neutrality level \cite{Localized}. In order to characterize further how the localized interface states contribute to the IPE yield, we measured the IPE yield spectra under applied reverse bias varying from 0.001 V to 5.00 V, as shown in Figure~\ref{fig:s1}. The humps around 1.33 eV are found for all metal electrodes although there is a certain degree of metal dependence in the relative strength of hump. It is readily noticeable that the hump become more pronounced with the applied bias increasing. This indicates that more and more electrons excited from the localized interface states can escape and transmit into the conduction band minima of GaAs with the aid of the electric field in the depletion region due to the applied bias.

\section{Finite Element Electrostatic Modeling}
~~
Finite element electrostatic modeling has been performed to find the electron energy
band alignment of metal/graphene/GaAs junction for high and low interface-trap density regions with a commercial software package FlexPDE \cite{PDE1,PDE2,PDE3}.

The electrostatic potential drop $U$ from the vacuum level on the metal electrode side was calculated by solving the Poisson's equation $\nabla^{2}U(x) = -\frac{\rho(x)}{\varepsilon_{0}\varepsilon_{r}}$, where $\rho(x)$ is the net charge density, $\varepsilon_{0}$ is the permittivity of vacuum, and $\varepsilon_{r}$ is the dielectric constant. The net charge density is assumed to be zero in the vacuum gaps (metal/graphene or graphene/GaAs contact) and is given as $\rho(x) = q[N_{D} - n(x)]$ in the GaAs substrate, where $q$ is the electronic charge, $N_{D} = 5\times10^{16}~cm^{-3}$ is the doping concentration of n-GaAs, $n = N_{C}\exp(-\frac{E_{C}-E_{F}}{k_{B}T}) $ is the concentration of free electron in GaAs, $N_{C}$ is the effective density of states in the conduction band of GaAs, $E_{C}(x) = \phi_{M} - \chi_{S} - qU(x)$ is the conduction band edge of GaAs, $E_{F}$ is the Fermi-level, $k_{B}T$ is the thermal energy, $\phi_{M}$ is the metal work-function, and $\chi_{S}$ is the electron affinity of GaAs.

We define the metal surface to be at $x=0$, the vacuum gap between metal surface and graphene layer in $0<x<d$, the graphene layer at $x=d$, the vacuum gap between graphene layer and GaAs substrate in $d<x<2d$, and the GaAs substrate in $2d<x<X$ as shown in Figure~S2. It is considered that the thin charge sheets $Q_{M}$, $Q_{G}$, and $Q_{SS}$ are located at $x=0$, $x=d$, and $x=2d$ \cite{InterfaceGraphene5,SzeText,SzePaper,TungPaper}, where $Q_{M}$ is the surface free charge density on the metal surface, $Q_{G} = {q\Delta E_{F} \vert \Delta E_{F} \vert \over \pi \hbar^2 v_{F}^2}$ is the doping charge density of graphene, $Q_{SS} = -qD_{it}(E_{g}-E_{CNL}-\phi_{B})$ is the interface-trap charge density on GaAs surface, $\Delta$$E_{F} = \phi_{M} - \phi_{G} - qU(d)$ is the Fermi-level shift from Dirac point of graphene, $\hbar$ is the reduced Planck constant, $v_{F}$ is the Fermi velocity of graphene, $\phi_{G}$ is the graphene work-function, $D_{it}$ is the interface-trap density on GaAs surface, $E_{g}$ is the band gap of GaAs, $E_{CNL}$ is the charge neutrality level of GaAs determined from the I-V and IPE measurements in this work, and $\phi_{B} = \phi_{M} - \chi_{S} - qU(2d)$ is the Schottky barrier of metal/graphene/GaAs junction. We assume that two opposite sign of interaction dipole charges are located at $x=0$ and $x=d$ and $Q_{D}$ represents the interaction dipole charge density on the graphene side.

The boundary conditions are assigned as $U(0) = 0$, $ U(2d+X) =  \phi_{M} - \chi_{S} - k_{B}T\ln(\frac{N_{C}}{N_{D}})$, $\lim _{\epsilon\to0} \int _{d-\epsilon}^{d+\epsilon}\rho(x)\,dx = Q_{G} + Q_{D}$, and $\lim _{\epsilon\to0} \int _{2d-\epsilon}^{2d+\epsilon}\rho(x)\,dx = Q_{SS}$. From the modeling, the Schottky barrier $\phi_{B}$, the voltage drop across metal/graphene interface $\Delta _{MG} = {(Q_{M}-Q_{D}) \over C}$, the voltage drop across graphene/GaAs interface $\Delta _{GS} = -{(Q_{SS}+Q_{SC}) \over C}$, and the Fermi-level shift $\Delta$$E_{F}$ of graphene can be obtained \cite{InterfaceGraphene5,SzeText,SzePaper,TungPaper}, where $C = {\varepsilon_0 \over d}$ is the capacitance of vacuum gap per unit area, $Q _{SC} = \sqrt {2\varepsilon_{0} \varepsilon_{r} N_{D} (\phi_{bi} - k_{B}T)}$ is the space charge density in GaAs, and $\phi_{bi} = \phi_{B} - k_{B}T\ln(\frac{N_{C}}{N_{D}})$ is the built-in potential energy. The sign of $\Delta$$E_{F}$ is determined by the charge carrier type of graphene, that is, $\Delta$$E_{F} > 0$ for p-type graphene and $\Delta$$E_{F} < 0$ for n-type graphene. Note that the depletion region completely ends way before the edge of calculation domain ($x = 2d + X$) with the thickness of 500 nm for the GaAs substrate so that no difference in calculation results is made from using the actual thickness of GaAs substrate (0.5 mm).

\begin{figure*}[!]
\sfigtwo
\includegraphics{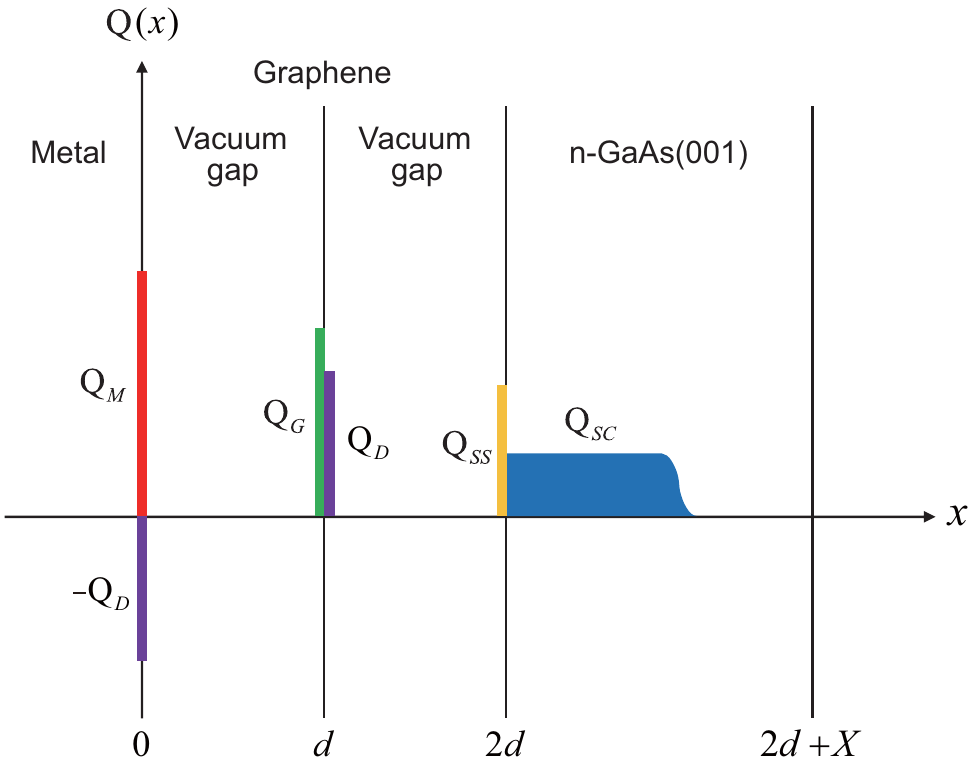}
\caption{\label{fig:s2}Charge distribution across metal/graphene/GaAs junction, where $Q_{M}$ is the surface free charge density on metal, $Q_{D}$ is the interaction dipole charge density at metal/graphene contact, $Q_{G}$ is the doping charge density of graphene, $Q_{SS}$ is the interface-trap charge density on GaAs surface, $Q_{SC}$ is the space charge density in the depletion region of GaAs substrate, $d$ is the thickness of vacuum gap, and $X$ is the thickness of GaAs substrate.}
\end{figure*}

\begin{table}[!]
\sfigone
\centering
\caption{The calculated $\phi_{B}$ and pinning strength $S$ by assuming $Q_{D}$ = 0 and varying $D_{it}$ for the metal/graphene/GaAs junctions with Al, Ti, Ni, and Pt electrodes.}
\resizebox{\textwidth}{!}{
\begin{tabular}{|c|c|c|c|c|c|c|c|c|c|}
\hline
\multicolumn{2}{|c|}{$D_{it}$ (eV$^{-1}$$\cdot$cm$^{-2}$)} & 0 & 1$\times 10^{12}$ & 5$\times 10^{12}$ & 1$\times 10^{13}$ & 5$\times 10^{13}$ & 1$\times 10^{14}$ & 5$\times 10^{14}$ & 1$\times 10^{15}$ \\ \hline
\multirow{4}{*}{\begin{tabular}[c]{@{}c@{}}Calculated $\phi_{B}$ (eV)\\(assuming $Q_{D}$ = 0)
\end{tabular}} 
 & Al & 0.187 & 0.229 & 0.356 & 0.459 & 0.704 & 0.766 & 0.826 & 0.834 \\ \cline{2-10} 
 & Ti & 0.283 & 0.322 & 0.442 & 0.535 & 0.738 & 0.784 & 0.829 & 0.836 \\ \cline{2-10} 
 & Ni & 0.681 & 0.691 & 0.722 & 0.746 & 0.805 & 0.821 & 0.937 & 0.840 \\ \cline{2-10} 
 & Pt & 0.808 & 0.810 & 0.816  & 0.821 & 0.834 & 0.838 & 0.841 & 0.842 \\ \hline
\multicolumn{2}{|c|}{S} & 0.412 & 0.384 & 0.301 & 0.234 & 0.082 & 0.045 & 0.010 & 0.005 \\ \hline
\end{tabular}}
\end{table}

\begin{table}[!]
\sfigtwo
\centering
\caption{The estimated $Q_{D}$ to match with the I-V measured $\phi_{B}$ by varying $D_{it}$ for the metal/graphene/GaAs junctions with Al, Ti, Ni, and Pt electrodes.}
\resizebox{\textwidth}{!}{
\begin{tabular}{|c|c|c|c|c|c|c|c|c|c|}
\hline
\multicolumn{2}{|c|}{$D_{it}$ (eV$^{-1}$$\cdot$cm$^{-2}$)} & 0 & 1$\times 10^{12}$ & 5$\times 10^{12}$ & 1$\times 10^{13}$ & 5$\times 10^{13}$ & 1$\times 10^{14}$ & 5$\times 10^{14}$ & 1$\times 10^{15}$ \\ \hline
\multirow{4}{*}{\begin{tabular}[c]{@{}c@{}}Estimated $Q_{D}$\\ ($10^{-6}~$C$\cdot$cm$^{-2}$)\\(to fit with the I-V\\measured $\phi_{B}$)
\end{tabular}} 
 & Al & -5.87 & -5.92 & -6.13 & -6.39 & -8.62 & -11.84 & -54.48 & -149.98 \\ \cline{2-10} 
 & Ti & -5.23 & -5.29 & -5.52 & -5.81 & -8.35 & -12.05 & -62.69 & -178.49 \\ \cline{2-10} 
 & Ni & 1.60 & 1.74 & 2.27 & 3.09 & 18.61 & 60.31 & 1286.07 & 5048.71 \\ \cline{2-10} 
 & Pt & 3.35 & 3.51 & 4.29 & 5.58 & 28.57 & 89.08 & 1843.13 & 7210.81 \\ \hline
\end{tabular}}
\end{table}

\begin{table}[!]
\sfigthree
\centering
\caption{Calculated relevant potentials ($\phi_{B}$, $\Delta E_{F}$, $\Delta _{MG}$, $\Delta _{GS}$) and charges ($Q_{M}$, $Q_{G}$, $Q_{SS}$, $Q_{SC}$) for metal/graphene/GaAs junctions with Al, Ti, Ni, and Pt electrodes. The calculation has been performed for high $D_{it} = 5\times10^{14}$ eV$^{-1}$ cm$^{-2}$ and low $D_{it} = 5\times10^{12}$ eV$^{-1}$ cm$^{-2}$ with fixed $Q_{D}$ obtained for $D_{it} = 5\times10^{12}$ eV$^{-1}$ cm$^{-2}$ in Table~S2.}
\resizebox{\textwidth}{!}{
\begin{tabular}{|c|c|c|c|c|c|c|c|c|}
\hline
\textbf{Metal} & \multicolumn{2}{c|}{\textbf{Al}} & \multicolumn{2}{c|}{\textbf{Ti}} & \multicolumn{2}{c|}{\textbf{Ni}} & \multicolumn{2}{c|}{\textbf{Pt}} \\ \hline
$\phi_{M}$ (eV)\cite{CRC,MP,Work-function}& \multicolumn{2}{c|}{4.08} & \multicolumn{2}{c|}{4.33} & \multicolumn{2}{c|}{5.15} & \multicolumn{2}{c|}{5.65} \\ \hline
$Q_{D}$ ($10^{-6}~$C$\cdot$cm$^{-2}$) & \multicolumn{2}{c|}{-6.13} & \multicolumn{2}{c|}{-5.52} & \multicolumn{2}{c|}{2.27} & \multicolumn{2}{c|}{4.29} \\ \hline
$D_{it}$ (eV$^{-1}$$\cdot$cm$^{-2}$) & 5$\times 10^{12}$ & 5$\times 10^{14}$ & 5$\times 10^{12}$ & 5$\times 10^{14}$ & 5$\times 10^{12}$ & 5$\times 10^{14}$ & 5$\times 10^{12}$ & 5$\times 10^{14}$ \\ \hline
$E_{CNL}$ (eV) & 0.582 & 0.413 & 0.582 & 0.413 & 0.582 & 0.413 & 0.582 & 0.413 \\ \hline
$\phi_{B}$ (eV) & 0.895 & 1.008 & 0.901 & 1.008 & 0.465 & 0.995 & 0.392 & 0.993 \\ \hline
$\Delta E_{F}$ (eV) & 0.521 & 0.539 & 0.529 & 0.546 & -0.039 & 0.167 & -0.135 & 0.106 \\ \hline
$\Delta_{MG}$ (V) & 0.941 & 0.959 & 0.699 & 0.716 & -0.689 & -0.483 & -1.285 & -1.044 \\ \hline
$\Delta_{GS}$ (V) & -0.056 & 0.039 & -0.058 & 0.032 & 0.074 & 0.398 & 0.097 & 0.456 \\ \hline
$Q_{M}$ ($10^{-6}~$C$\cdot$cm$^{-2}$) & -3.364 & -3.306 & -3.467 & -3.412 & 0.234 & 0.848 & 0.498 & 1.215 \\ \hline
$Q_{G}$ ($10^{-6}~$C$\cdot$cm$^{-2}$) & 3.197 & 3.424 & 3.295 & 3.511 & -0.018 & 0.328 & -0.216 & 0.133 \\ \hline
$Q_{SS}$ ($10^{-6}~$C$\cdot$cm$^{-2}$) & 0.042 & -0.251 & 0.047 & -0.231 & -0.302 & -1.308 & -0.361 & -1.480 \\ \hline
$Q_{SC}$ ($10^{-6}~$C$\cdot$cm$^{-2}$) & 0.125 & 0.133 & 0.125 & 0.133 & 0.087 & 0.132 & 0.079 & 0.132 \\ \hline
\end{tabular}}
\end{table}

\section{Raman Spectrum Measurements}
\begin{figure*}
\sfigthree
\includegraphics{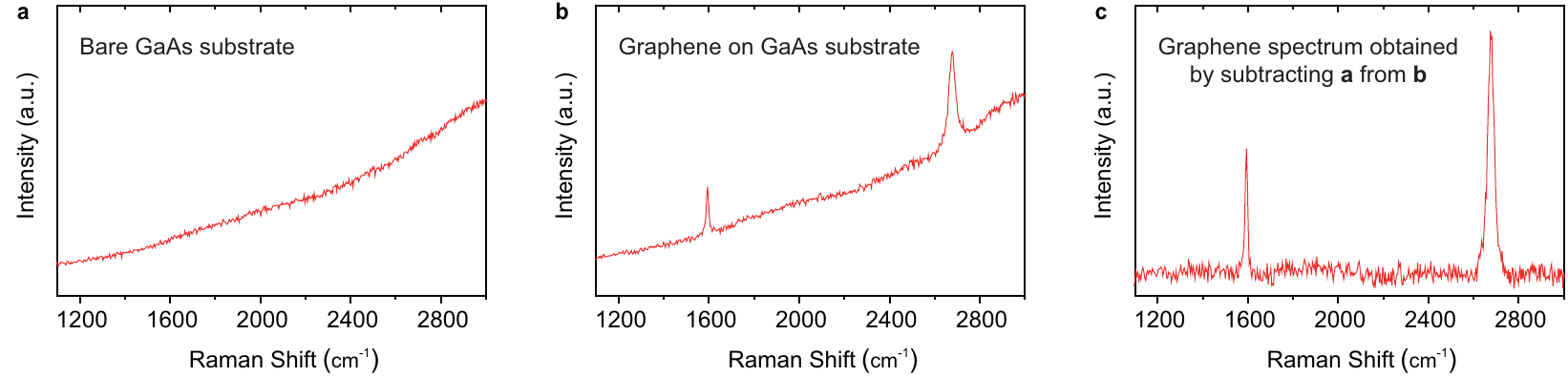}
\caption{\label{fig:s3}Raman spectrum measurements of the graphene used in this work. The Raman spectra of bare GaAs substrate (a), graphene on GaAs substrate (b), and graphene with the GaAs background signal subtracted (c).}
\end{figure*}

~~
The confocal Raman spectroscopy with the laser wavelength of 532 nm has been carried out for the transferred graphene on GaAs substrate. The measured spectra are shown in Figure~\ref{fig:s3}. The intensity ratio between the G peak ($\sim$1580 cm$^{-1}$) and the 2D peak ($\sim$2700 cm$^{-1}$) indicates that the transferred graphene is a monolayer. \cite{Raman}.

~

~

~

~

~

~

~

~

~

~

~

~

~

~

\section{Transmission Electron Microscope Images}
\begin{figure*}
\sfigfour
\includegraphics{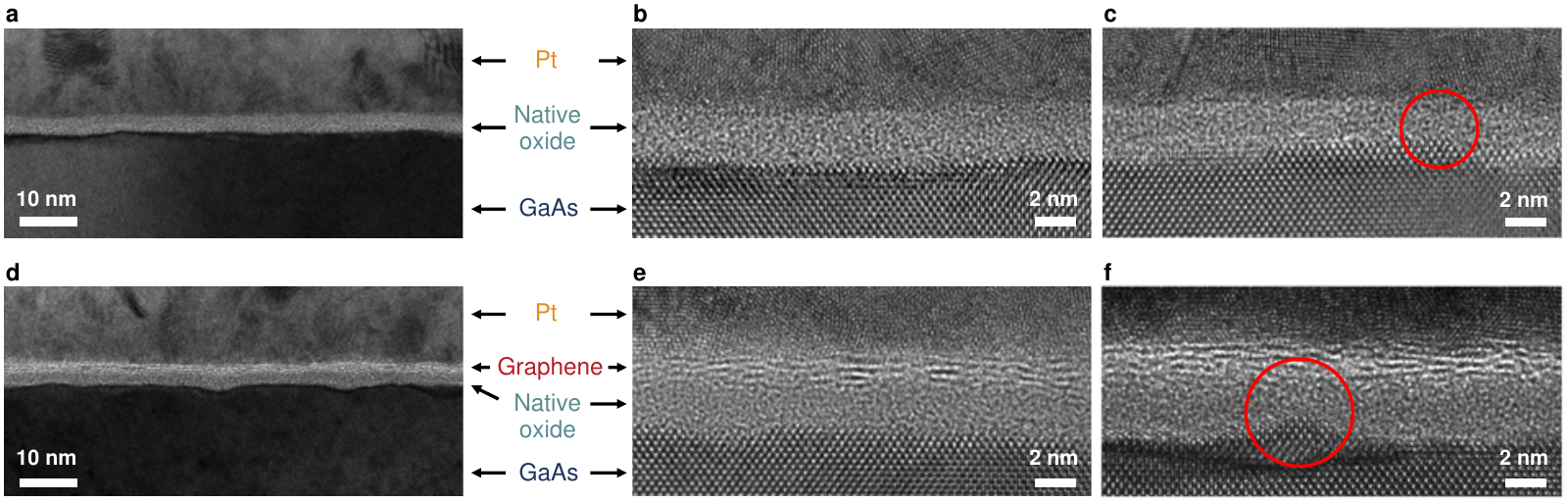}
\caption{\label{fig:s4}BFTEM (a) and HRTEM (b, c) images of Pt/GaAs junction. BFTEM (d) and HRTEM (e, f) images of Pt/graphene/GaAs junction. The red circles represent the small-size regions with very thin native oxide layer.}
\end{figure*}

~~
After all the electrical measurements were completed, transmission electron microscope images were obtained. The bright-field TEM (BFTEM) and high-resolution TEM (HRTEM) images of Pt/GaAs and Pt/graphene/GaAs junctions are shown in Figure~\ref{fig:s4}. The graphene layer inserted at Pt/GaAs interface is seen clearly \cite{TEM}. It is confirmed with HRTEM images taken on the several different sites in the Pt/GaAs (Figure~\ref{fig:s4}b,c) and Pt/graphene/GaAs (Figure~\ref{fig:s4}e,f) junctions that the thickness of native oxide layer is spatially inhomogeneous. In addition, the area of GaAs surface occupied with very thin or no native oxide layer is found to be quite small, indicated by red circles. The thickness of oxide layer on GaAs surface is estimated about $\sim$2.82 nm for Pt/GaAs junction and $\sim$2.27 nm for Pt/graphene/GaAs junction \cite{Oxide1}. It is likely to have slightly thicker oxide layer for Pt/GaAs junction compared to that for Pt/graphene/GaAs junction on the same GaAs substrate because the oxide layer on the graphene-covered area was passivated while the oxide layer on the graphene-uncovered area was exposed to the air before the metal electrode deposition. Since the native GaAs oxide is found to have very small band gap \cite{Oxide2}, the charge carrier transport across metal/GaAs junction depends significantly on whether the graphene layer is inserted at the interface or not rather than the small difference in the thickness of GaAs oxide layer.

